\documentclass[iop]{emulateapj}

\def\msun{\hbox{{\it M}$_{\odot}$}}
\def\mdot{\hbox{$\dot {\it M}$}}
\def\micron{$\mu$m}
\def\microns{$\mu$m}
\def\lsun{\rm {\it L}_{\sun}}
\def\rsun{$\mbox{{\it R}}_{\odot}$}
\newcommand\msunyr{\rm {\it M}_{\odot}\,yr^{-1}}
\newcommand\be{\begin{equation}}
\newcommand\en{\end{equation}}

\usepackage{graphicx}
\usepackage[caption=false]{subfig}

\newcounter{column_number}
\begin{document}

\shortauthors{Espaillat et al.}
\shorttitle{An Incipient Debris Disk in Cha I}

\title{An Incipient Debris Disk in the Chamaeleon I Cloud}

\author{
C. C. Espaillat\altaffilmark{1}, 
{\'A}. Ribas\altaffilmark{1},
M. K. McClure\altaffilmark{2}, 
J. Hern{\'a}ndez\altaffilmark{3},
J. E. Owen\altaffilmark{4,5},
N. Avish\altaffilmark{1},
N. Calvet\altaffilmark{6},
\& R. Franco-Hern{\'a}ndez\altaffilmark{7}
}

\altaffiltext{1}{Department of Astronomy, Boston University, 725 Commonwealth Avenue, Boston, MA 02215, USA; cce@bu.edu, aribas@bu.edu, navish@bu.edu}
\altaffiltext{2}{Karl-Schwarzschild-Stra{\ss}e 2, 85748 Garching bei M{\"u}nchen; mmcclure@eso.org} 
\altaffiltext{3}{Instituto de Astronom{\'i}a, Universidad Aut—noma Nacional de M{\'e}xico, Campus Ensenada; MX; hernandj@astrosen.unam.mx} 
\altaffiltext{4}{Institute for Advanced Study, Einstein Drive, Princeton, NJ 08540, USA; jowen@ias.edu}
\altaffiltext{5}{Hubble Fellow}
\altaffiltext{6}{Department of
Astronomy, University of Michigan, 830 Dennison Building, 500 Church
Street, Ann Arbor, MI 48109, USA; ncalvet@umich.edu}
\altaffiltext{7}{Instituto de Astronom{\'i}a y Meteorolog{\'i}a, Universidad de Guadalajara, Avenida Vallarta No. 2602, Col. Arcos Vallarta, CP 44130, Guadalajara, Jalisco, M{\'e}xico; rfranco@astro.iam.udg.mx} 

\begin{abstract} 

The point at which a protoplanetary disk becomes a debris disk is difficult to identify.  To better understand this, here we study the $\sim$40~AU separation binary T~54 in the Chamaeleon I cloud.  We derive a K5 spectral type for T~54~A (which dominates the emission of the system) and an age of $\sim$2~Myr.  However, the dust disk properties of T~54 are consistent with those of debris disks seen around older and earlier-type stars.  At the same time, T~54 has evidence of gas remaining in the disk as indicated by [\ion{Ne}{2}],  [\ion{Ne}{3}], and [\ion{O}{1}] line detections. 
We model the spectral energy distribution of T~54 and estimate that  $\sim$3$\times$10$^{-3}$ M$_{\Earth}$ of small dust grains ($<$0.25~{\micron}) are present in an optically thin circumbinary disk along with at least  $\sim$3$\times$10$^{-7}$ M$_{\Earth}$ of larger ($>$10~{\micron}) grains within a circumprimary disk.  
Assuming a solar-like mixture, we use Ne line luminosities to place a minimum limit on the gas mass of the disk ($\sim$3$\times$10$^{-4}$ M$_{\Earth}$) and derive a gas-to-dust mass ratio of $\sim$0.1.
We do not detect substantial accretion, but
we do see H$\alpha$ in emission in one epoch, suggestive that there may be intermittent dumping of small amounts of matter onto the star. 
Considering the low dust mass, the presence of gas, and young age of T~54, we conclude that 
this system is on the bridge between the protoplanetary and debris disk stages.

\end{abstract}


\section{INTRODUCTION} \label{intro} 

Protoplanetary disks provide the reservoirs of gas and dust needed to form the multitude of planets that have been detected to date \citep[e.g.,][]{fischer14}.  Accordingly, the lifetimes of protoplanetary disks set the timescale for planet formation.  
The dispersal of disks around pre-main sequence (PMS) stars is dominated by accretion and likely also photoevaporation by high-energy stellar radiation \citep{hollenbach94, clarke01, alexander06, gorti09a, alexander14} and to some degree the planet formation process \citep[see][and references within]{espaillat14}.

Observations have shown that the fraction of PMS stars displaying near-infrared (NIR) emission within a star-forming region drops to less than 10$\%$ by 5 Myr \citep[e.g.,][]{haisch01b, hernandez07a, ribas15}.  
\citet{fedele10} suggest that the accretion rates onto young stars decrease on a slightly shorter timescale.  However, these studies are only sensitive to material very close to the star.  NIR emission traces the dust in the innermost $\sim$AU of the disk and an accretion rate measures gas on or close to the stellar surface.  
Such information does not provide information on how the spatial distribution of disk material several AU from the star changes with age.  At the millimeter wavelengths, it is possible to trace large grains throughout the disk since disks are mostly optically thin in this wavelength regime.  \citet{andrews10} find marginal evidence of evolution of the millimeter emission of disks between the 1 Myr old Taurus region \citep{kh95} and the older $\sim$10 Myr old Upper Sco region \citep{pecaut12}.
Therefore, the most we can conclude is that the timescale of dissipation of the innermost disk is a few Myr.  More work needs to be done to understand the dissipation of the outer disk.

We can approach the issue of narrowing down the timescale for disk dissipation of the outer disk from a different angle by studying protoplanetary disks at the end of their primordial lifetime, before becoming a debris disk. This presumably transient stage is difficult to define and identify.

To explain observations to date, \citet{wyatt15} suggested that the following steps occur as a protoplanetary disk evolves into a debris disk, with the caveat that the order of these stages is not certain.  First, there is the transitional disk stage when there is a significant depletion of material in the inner disk, but there is still a substantial outer disk composed of gas and dust \citep[e.g.,][]{espaillat14}. Depletion of mm-sized dust in the outer disk occurs \citep[e.g.,][]{testi14} along with the creation of hot second-generation dust in the inner disk regions.  Lastly, gas is removed \citep[e.g.,][]{alexander14} and ringed concentrations of planetesimals form.

As noted by \citet{wyatt15}, there is no formal distinction between protoplanetary and debris disks, but there are a few distinctions that have been invoked previously.   The mass of protoplanetary disks is dominated by gas while the mass of debris disks is thought to be dominated by dust.  Typically, an age of 10~Myr is used to separate the two stages.  
Although, \citet{hernandez09} find that for more massive stars the debris disk phenomenon can start much earlier (i.e., 5~Myr).
Protoplanetary disks are mostly optically thick to the stellar radiation while debris disks are optically thin.  Also, the material in protoplanetary disks is considered primordial (i.e., inherited from the natal molecular cloud) while the material in debris disks is thought to be second-generation (i.e., created by collisions of planetesimals).

In recent years, the boundary between protoplanetary and debris disks has been further muddled. Some gas has been detected in about a dozen debris disks, mainly surrounding young ($<$50~Myr) A-type stars \citep{kospal16} with gas-to-dust ratios of about 1 \citep[e.g.,][]{zuckerman95,lagrange98,roberge06,redfield07,maness08,moor11}. 
In particular, the CO gas line has been detected in the millimeter from
the 5 Myr old HD~141569 \citep{white16},
the 12 Myr old Beta Pic \citep{dent14},
the 16 Myr old HD~131835 \citep{moor15},
the 30 Myr old HD~21997 \citep{moor11}, and
the 40 Myr old 49~Cet \citep{hughes08}.
Most of these CO detections have been attributed to the second-generation process of CO outgassing via collisional cascades involving icy exocomets planetesimals \citep{zuckerman12, kral17}. This is the most likely scenario in cases where the gas is coincident with the dust in the disk, as is seen in most of the systems outlined above. 
Alternatively, the gas may have primordial origins \citep{matthews14}.
For example, resolved millimeter images of HD~21997 reveal that the gas is located closer to the star than the dust, suggestive that the gas has a primordial origin \citep{moor13}.

Disks on the bridge between the protoplanetary and debris disk stages can be used to better understand both disk dissipation processes and timescales, clarifying the distinction between these two stages.
Presumably, such disks would have mostly dissipated their primordial material and hence be less massive with accordingly weak IR emission and slow or non-detectable accretion onto the star, and studies have revealed such objects \citep{cieza07, luhman10, williams11}.  
\citet{hardy15} used ALMA to study a subset of these objects, namely low-mass PMS (i.e., T Tauri stars; TTS) that displayed weak mid-infrared (MIR) and far-infrared (FIR) excesses. The sample consisted of TTS that had no measurable accretion onto the star (i.e., weak TTS; WTTS).  In contrast, classical TTS (CTTS) do have ongoing accretion onto the star and are
typically associated with the presence of substantial circumstellar material, which leads to accretion onto the star and significant IR excess \citep[see][and references within]{hartmann16}.  WTTS are often diskless, with no IR excess and no accretion onto the star. However, there are exceptions.  As mentioned above, there are some objects with IR excesses, but no measurable accretion onto the star.  One must keep in mind though that an object may be classified as a WTTS, but it may still have accretion onto the star at a slow rate that cannot be measured with typical methods \citep{ingleby11b}.
In their sample of 24 WTTS with weak IR excess,
 \citet{hardy15} detected no CO gas line emission and only 4 objects were detected in the dust continuum at 1.3~mm.  The upper limits to the dust and gas masses suggest that many objects in the sample may be young debris disks.

Here we study in more detail one of the objects in \citet{hardy15} that had no CO gas line nor dust continuum detection: the T~54 system in Chamaeleon I.  This object was first identified by \citet{henize73} and labeled HM Anon.  It is also known as J11124268-7722230 (2MASS), CHX22 \citep{feigelson89}, 11111-7705 (IRAS), and Ass Cha T2-54 (SIMBAD). T~54 is a $\sim$0.2$^\prime$$^\prime$ separation binary, which corresponds to about 40 AU at 160 pc \citep{lafreniere08,daemgen13}. T~54~A dominates the emission of the system.
\citet{lafreniere08} and \citet{daemgen13} report $\Delta J$, $\Delta H$, and $\Delta K$ for T~54 and find that the primary is about 4--5 times brighter than the secondary at each of these bands.  
{\it Spitzer} IRS spectra of this system display little NIR excess at short wavelengths and a rise at longer wavelengths \citep{kim09}, seemingly consistent with a cleared inner disk due to dynamical clearing by the companion \citep{lubow06}.   
Circumbinary disks can offer interesting insights to disk dispersal. First, the amount of dynamical clearing due to the companion in the disk is roughly constrained \citep{lubow06}. Second, circumbinary disk evolution should proceed at a more accelerated pace since there is emission from two stars impinging on the disk and possibly enhancing photoevaporation \citep{alexander12}. 

In this work, we measure the dust and gas properties of the T~54 system.   We explore the gas in the disk by obtaining accretion rate indicators with high-resolution optical and near-infrared (NIR) spectra with {\it Magellan} and analyzing archival HST ACS spectra ({\S}~2).
In {\S}~3, we study the dust component of the disk by modeling its spectral energy distribution (SED).  In {\S}~4, we explore whether T~54 is better categorized as a protoplanetary or debris disk. We end with our summary and conclusions in {\S}~5.

\section{OBSERVATIONS \& DATA REDUCTION} \label{redux}

\subsection{FIRE}

We performed NIR echelle spectroscopy on T~54 as well as CVSO~207 and CR~Cha, which we use as comparison stars, from 2011 July 18 through 2011 July 19 with the Folded-port InfraRed Echellette (FIRE) spectrometer \citep{simcoe08} 
on the 6.5~m Magellan Baade Telescope at Las Campanas Observatory.  
FIRE provides spectral coverage from 0.8--2.5~{\micron}.  We used the 0.45$^\prime$$^\prime$ slit to acquire all of the light from the objects.  The resulting spectral resolution was $\lambda/\delta\lambda$ = 8,000.  Each target was placed on the slit using the $J$-band camera and then dithered in an ABBA pattern.  The `high gain' and `sample up the ramp' modes were used during readout. The integration times for T~54 and the comparison stars were about 100~s and 20~s, respectively.
A0 telluric standard stars were observed at an airmass appropriate for each target observation.  Wavelength calibrations were obtained using exposures of the ThAr lamp, while we used the high voltage quartz lamp on the flat-field screen to map the pixel response.  To reduce the data, we used the FIREhose instrument pipeline, with default settings, as described in \citet{bochanski11}.

\subsection{MIKE} 

We obtained optical {\it Magellan} Inamori Kyocera Echelle (MIKE) double-echelle spectrograph \citep{bernstein03} observations of T~54 on
the 6.5~m {\it Magellan Clay Telescope} at Las Campanas Observatory on 2007 Feb 10, 2008 Feb 15, 2008 Feb 18, and 2009 Jan 18. We also obtained data of CS~Cha, which we use in our analysis as a comparison star, on 2009 Jan 18.  
MIKE spectra span 4800--9000~{\AA} and we used a slit size of
0.7$^\prime$$^\prime$$\times$5.0$^\prime$$^\prime$ (R$\sim$35,000) and 2$\times$2 pixel on-chip binning with the fast readout mode.  
Exposure times ranged  between 150--240~s. 
As above, we used the ThAr lamp and the high voltage quartz lamp to perform calibrations.  
Data were reduced
using the MIKE data reduction
pipeline.\footnote{http:$/$$/$web.mit.edu$/$$\sim$burles$/$www$/$MIKE$/$} 

\section{Analysis \& Results}

In the following section, we use new high-resolution optical spectra of T~54 to measure a spectral type and derive new stellar parameters for the primary star, which dominates the emission of the system.  We investigate multiple gas indicators: archival spectra tracing H$_{2}$ and new data of the Br$\gamma$, \ion{He}{1}, and H$\alpha$ line regions.  Lastly, we model the system using an optically thin dust model adopting the stellar parameters derived here.

\subsection{Stellar Properties} \label{star}

\subsubsection{Spectral Type} \label{spt}

\begin{figure*}[ht!]
\epsscale{1.0}
\plotone{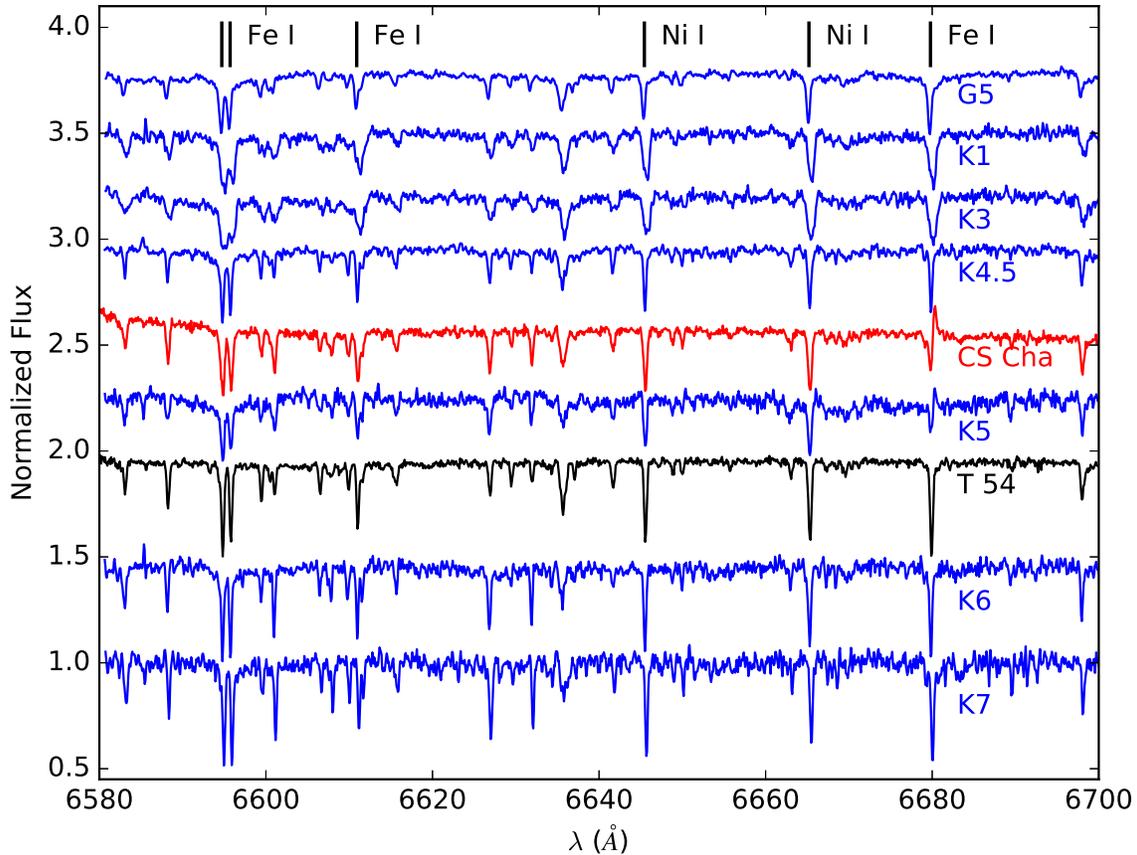}
\caption[]{
Optical spectra of T~54 (black) and CS~Cha (red) compared to the following spectral standard stars (blue) from \citet{hernandez14}: 
SO~243 (G5),
SO~791 (K1),
SO~615 (K3),
SO~583 (K4.5),
SO~1156 (K5),
SO~697 (K6),
SO~929 (K7).
Some photospheric lines are denoted and were identified using the revised version of the ILLSS Catalogue \citep{coluzzi99}. We note the presence of the \ion{He}{1} emission line at $\sim$6680~{\AA} in the CS~Cha and the K5 template spectra; both objects are CTTS and \ion{He}{1} can be seen in emission in fast accretors. We measure a spectral type of K5${\pm}$1 for both T~54 and CS Cha. 
}
\label{figspt}
\end{figure*}

Here we utilize our high-resolution optical spectra to derive spectral types for our target T~54~A as well as CS~Cha, which we will use in later analysis as a template for a CTTS.  For our spectral standards, we use objects from \citet{hernandez14} who determined spectral types for a large sample of stars in the $\sigma$ Orionis cluster. 
These spectra were obtained with HECTOSCHELLE (R$\sim$34,000), which has a spectral resolution similar to that of our MIKE spectra (R$\sim$35,000).

We measured spectral types for T~54~A and CS~Cha via direct comparison with the spectroscopic standard stars (Figure~\ref{figspt}) by minimizing the differences (root mean squared) after continuum normalization between these stars and the templates.
We derive K5$\pm$1 for both targets.  
Our derived spectral type for T~54~A differs from previous literature measurements.
\citet{manara16} measure a spectral type of K0 for this object, which is consistent within the uncertainties with the spectral type of G8 reported by \citet{luhman07} and \citet{daemgen13}.  
Our derived spectral type for CS~Cha is consistent with \citet{luhman04} within the uncertainties, but again differs from \citet{manara14} who report a spectral type of K2 for this object.    
While we do not have a G8 or K0 spectral standard star, it is clear in Figure~\ref{figspt} that T~54 does not resemble the G5 or K1 template star; its absorption line depths are between those seen the K5 and K6 templates.  CS~Cha also more closely resembles the later K-type stars than the earlier K-type stars; its absorption line depths are similar to the K4.5 and K5 templates.

One possible explanation for the discrepancies between our measured spectral types and previous work may be the spectral resolutions of the data.  Our data have a resolution of $\sim$35,000 while \citet{manara16} and \citet{manara14} use X-shooter spectra with a resolution of $\sim$17,000. The spectra used by \citet{daemgen13}, \citet{luhman07}, and \citet{luhman04} all had resolutions of $\sim$1,500.  \citet{manara14} also note that their early K-type stars have larger uncertainties since their photospheric templates are incomplete in late G-type and early K-type stars.  

\subsubsection{Extinction, Mass, Luminosity, \& Age} \label{star}

Our derived stellar parameters for T~54~A are listed in Table~1.
As mentioned in $\S$~1, the emission observed from the T~54 system is dominated by T~54~A, which is about 4--5 times brighter than the secondary in the NIR \citep{lafreniere08, daemgen13}.   
To faciliate comparison with previous work on T~54, below we adopt colors and effective temperatures from \citet{kh95}.  \citet{pecaut13} report slightly different colors and an effective temperature for a K5 star of 4150~K (a 5$\%$ difference), which would not lead to significant variations in calculated parameters. We measure a 
visual extinction (A$_{V}$) of 0.2 using the extinction law of \citet{mathis90} with an R$_{V}$ of 3.1.  
Extinctions were calculated by comparing V-R, V-I, and R-I colors from \citet{rydgren80} to the
photospheric colors of a K5 star from \citet{kh95}.   
We derive a stellar mass (M$_{*}$) of 1.1 ${\msun}$
from the HR diagram
and the \citet{siess00} evolutionary tracks using the stellar temperature (T$_{*}$) and luminosity (L$_{*}$).
We adopt a stellar temperature of 4350~K from \citet{kh95} based upon the K5 spectral type derived here. 
The luminosity of T~54~A (1.9 ${\lsun}$) was calculated following \citet{kh95} with dereddened J-band photometry \citep[2MASS;][]{cutri03} and a
distance of 160~pc $\pm$ 15~pc \citep{whittet97}. 
From the derived luminosity and adopted temperature, we measure R$_{*}$ of 2.5 {\rsun}. Using the \citet{siess00} evolutionary tracks we find an age of 1.8 Myr for T~54 (assuming that the secondary and primary are the same age), which is consistent with the age of the Cha I cloud of about 1 -- 6 Myr \citep{luhman07}. Considering that up to 20$\%$ of the total luminosity of T~54 may be due to T~54~B would lead to a derived age of 2.3~Myr.  Taking into account the uncertainties due to the distance we arrive at ages of 1.4 -- 2.3~Myr. We also note that in general age estimates of individual young stars are highly uncertain \citep{soderblom14, herczeg15}.  

\begin{deluxetable}{l c c c}
\tabletypesize{\scriptsize}
\tablewidth{0pt}
\tablecaption{Stellar Properties}
\startdata
\hline
\colhead{Property} & \colhead{Value}\\
\hline
M$_{*}$ (M$_{\sun}$) & 1.1  \\
R$_{*}$ (R$_{\sun}$) & 2.5  \\
T$_{*}$ (K) & 4350 \\
L$_{*}$ ($\lsun$) & 1.9  \\
A$_{V}$ & 0.2   \\
Spectral Type & K5  \\
Age (Myr) & 1.8
\enddata
\end{deluxetable}

\subsection{Accretion Properties}

T~54 has been classified by many previous studies as a non-accreting object.
\citet{nguyen12} 
observed H$\alpha$ in absorption \citep{white03}.  
\citet{daemgen13} spatially resolved T~54~A and T~54~B and measured upper limits to the accretion using the Br$\gamma$ line following the correlation found by \citet{muzerolle98}. T~54~A and T~54~B have upper limits to their Br$\gamma$ line luminosities corresponding to accretion rates of $<$9.6$\times$10$^{-10}$ $\msunyr$ and $<$3.5$\times$10$^{-10}$ $\msunyr$.
Using high-resolution spectra of T~54, \citet{manara16} find an upper limit of about 3$\times$10$^{-10}$$\msunyr$ using the Br$\gamma$ line.
Below we study the H$_{2}$, Br$\gamma$, \ion{He}{1}, and H$\alpha$ line regions in T~54 to search for signs of gas accretion in this system.

\subsubsection{H$_{2}$}

\begin{figure}
\epsscale{1.18}
\plotone{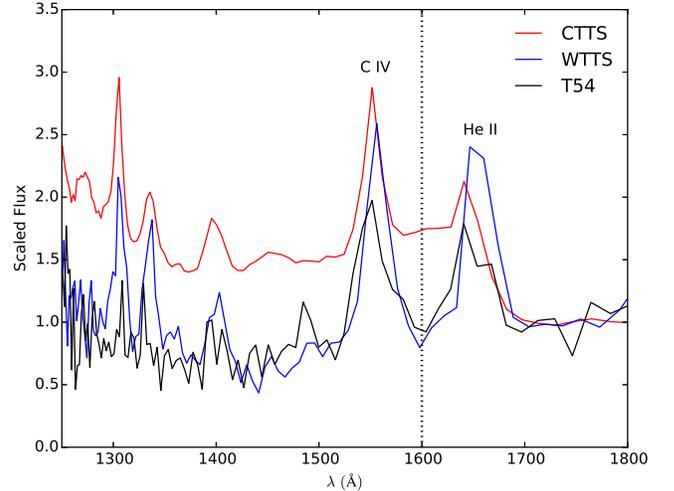}
\caption[]{
HST ACS FUV spectra of T~54 (black), a CTTS (red, CS Cha), and a K3 WTTS (blue, MML~28) from \citet{ingleby09}.  To enable comparison of the 1600~{\AA} H$_{2}$ feature (dotted line), spectra are scaled to the continuum between 1700--1800~{\AA}, where there is no H$_{2}$ emission.
}
\label{figacs}
\end{figure}

We revisit archival HST FUV data of T~54 previously published by \citet{ingleby09}.
In Figure~\ref{figacs}, we compare T~54 to a CTTS (CS Cha) and a WTTS (MML~28). CS~Cha is a K5 star ({\S}~\ref{spt}) and MML~28 is a K3 star \citep{torres06}.
The CTTS spectrum is more filled in than that of T~54 at 1600~{\AA} between the \ion{C}{4} and \ion{He}{2} lines.  This ``bump'' at 1600~{\AA} 
is mostly due to H$_{2}$ excited by electron collisions \citep{bergin04, ingleby09}.
In contrast, T~54's emission in this region resembles the WTTS more closely.
\citet{ingleby09} found that in WTTS, the ratio of \ion{He}{2} to \ion{C}{4} is close to 1; for CTTS, it is less than 1.  For T~54, the ratio of \ion{He}{2} to \ion{C}{4} is about 1 as well.  Also, the slope of the CTTS at the shorter wavelengths ($<$1500~{\AA}) is rising, which is typical of CTTS \citep{ingleby09}, while T~54's slope is not rising, as is seen in the WTTS.  Based on this FUV data, it appears that T~54 is most similar to a WTTS, indicating that at the time of these observations there was no gas close to the star.

\subsubsection{Br$\gamma$ \& \ion{He}{1}}

\begin{figure}
\epsscale{1.18}
\plotone{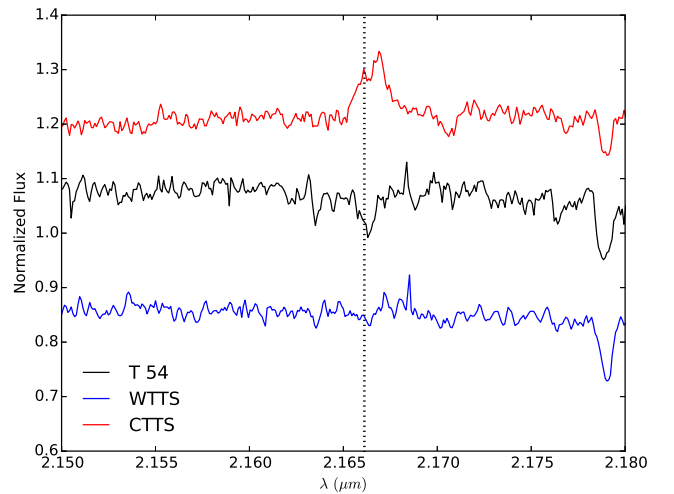}
\caption[]{
FIRE NIR spectra of T~54 (black), a K4 WTTS (blue: CVSO~207), and a K4 CTTS (red: CR~Cha).  The Br$\gamma$ wavelength is denoted with the vertical dotted line.  There is no discernible Br$\gamma$ emission in T~54. 
}
\label{figfire}
\end{figure}

\begin{figure}
\epsscale{1.18}
\plotone{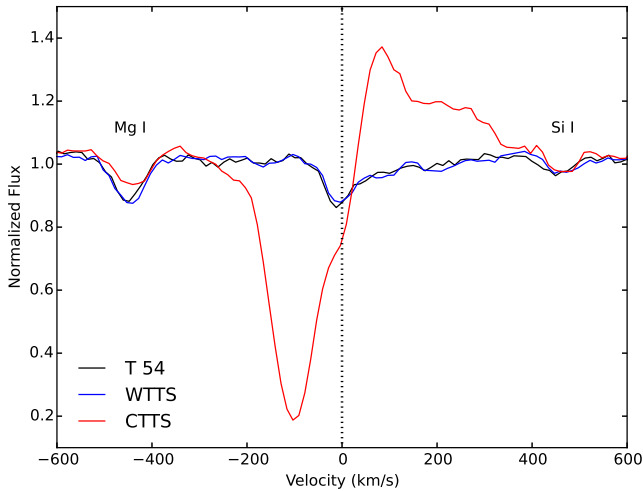}
\caption[]{
Profiles of the \ion{He}{1} ${\lambda}$10830 line region (dotted line) in T~54 (black), a WTTS template (blue, CVSO~207), and a CTTS (red, CR~Cha).  There is no evidence in T~54 for red-shifted absorption, as seen by \citet{ingleby11b} in slow accretors. 
}
\label{figfirehe}
\end{figure}

The FIRE spectra of T~54 show no detectable 
Br$\gamma$ emission (Figure~\ref{figfire}). For comparison we include a K4 WTTS \citep[CVSO~207;][]{briceno07} and a K4 CTTS \citep[CR~Cha;][]{torres06}. We note that the dip at the Br$\gamma$ line wavelength in the T~54 data is a residual artifact of the telluric correction. There is also no observed red-shifted absorption in the \ion{He}{1} (${\lambda}$10830) line in T~54 (Figure~\ref{figfirehe}).  \citet{ingleby11b} found that this line shows red-shifted absorption even in slow accretors of $\sim$10$^{-10} {\msunyr}$.  Based on our infrared spectra, we conclude that there was no measurable accretion in T~54 at the time of these observations.

\subsubsection{H$\alpha$}  \label{sec:halpha}

We also studied the region around the H$\alpha$ line in the T~54 system using MIKE high-resolution optical spectra.  This line is frequently
used in the literature to trace accretion onto stars via its equivalent width and the velocity width of the line 
\citep{white03,barrado03,natta04}. 
WTTS have an H$\alpha$ emission profile that is thought to be composed of photospheric absorption and chromospheric and coronal emission.  Hence, the H$\alpha$ emission profile of a WTTS is relatively narrow when found in emission.  In CTTS, the H$\alpha$ line is detected in emission and it is much wider due to an extra component from the accretion flow onto star which broadens the line \citep[e.g.,][]{hartmann16}.

\begin{figure}
\epsscale{1.18}
\plotone{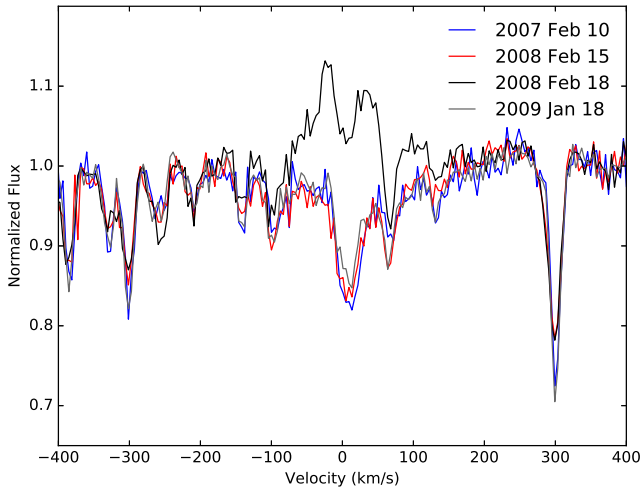}
\caption[]{
High-resolution MIKE spectra of T~54 in the H$\alpha$ ${\lambda}$6563 line region.  The spectra
have been normalized by the continuum for clarity and are colored according to date as noted in the figure key.  T~54 displays an H$\alpha$ emission feature on 2008 Feb 18 (black).  Note that three days earlier (red) this feature was not present.  
}
\label{figmike}
\end{figure}

In Figure~\ref{figmike}, we show four epochs of MIKE data.  In one epoch (2008 Feb 18), there is an H$\alpha$ emission line, while in the other epochs there is an H$\alpha$ absorption line. Interestingly, the observation with the H$\alpha$ emission (2008 Feb 18) was taken only 3 days after an observation with no H$\alpha$ emission (2008 Feb 15).

\begin{figure}
\subfloat{
\includegraphics[width=\linewidth]{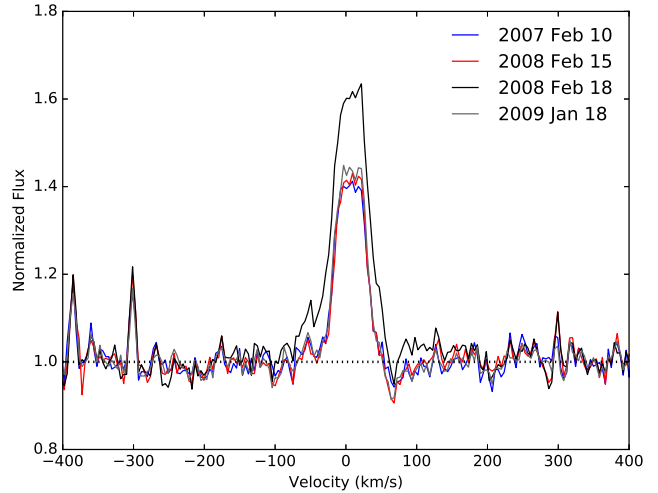}
}

\subfloat{
\includegraphics[width=\linewidth]{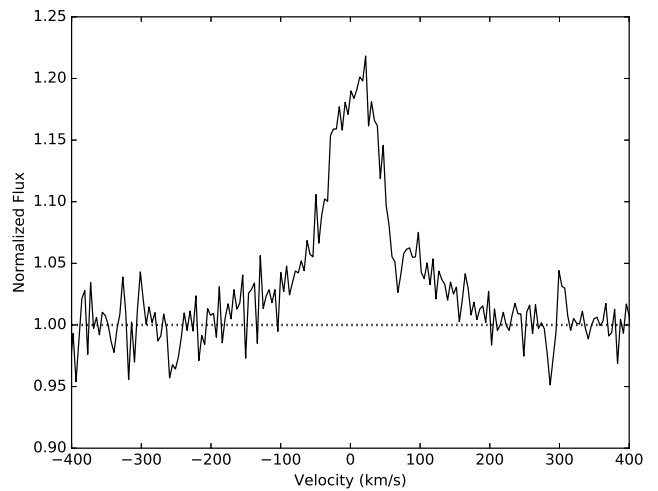}
}
\caption[]{
Residuals of the H$\alpha$ profiles presented in Figure~\ref{figmike}.  Top: We show residuals after subtracting a K5 field star from each epoch. Bottom: Residual H$\alpha$ profile for the 2008 Feb 18 epoch after subtracting the mean spectrum of the other three epochs.
}
\label{figresiduals}
\end{figure}

To further investigate this, we calculated field-star-subtracted H$\alpha$ line profile residuals (Figure~\ref{figresiduals}, top).  We used HECTOSCHELLE spectra of the field star SO~903, a K5.5 star with no \ion{Li}{1} nor H$\alpha$ emission from \citet{hernandez04}.  The residuals are fairly symmetric.  We note that in the H$\alpha$ emission epoch (black, Figure~\ref{figresiduals}, top) the emission line is somewhat stronger and wider.  We measure equivalent widths (EW) of 1~{\AA} for the H$\alpha$ emission epoch (2008 Feb 18) and about 0.6~{\AA} for the other three epochs. Since there are hints of wings in the H$\alpha$ profile residuals in the H$\alpha$ emission epoch (black, Figure~\ref{figresiduals}, top), we also calculate self-subtracted H$\alpha$ profile residuals by subtracting the mean profile of the other 3 epochs from the emission epoch (Figure~\ref{figresiduals}, bottom).  Here we see a narrow central component, but with faint traces of a broad component. We will return to this point in {\S}~\ref{sec:discusshalpha}.

\citet{white03} define a TTS as an accretor if the full width of the H$\alpha$ emission profile at 10$\%$ of the line peak is 270 km s$^{-1}$, independent of the spectral type.  Others have used a cut-off of 200 km s$^{-1}$ for very low-mass stars \citep{jayawardhana03}. Interestingly, we measure $\sim$200 km s$^{-1}$ for the self-subtracted H$\alpha$ profile residual of T~54.
To obtain a mass accretion rate estimate, we measured the width of
the self-subtracted H$\alpha$ emission line profile in the wings at 10\% of the maximum flux and, assuming that the emission is due to accretion, used the observed correlation between
line width and $\mdot$ \citep{natta04}.  We measure an accretion rate of $\sim$3$\times$10$^{-12}$ $\msunyr$, but note this is very uncertain given that the H$\alpha$ line is so weak.    
In addition, the relationship of \citet{natta04} was derived using stars with $>$200 km s$^{-1}$; T~54 is just in this limit. 
Based on this and the additional accretion indicators presented above, we conclude that T~54 is not actively accreting at a detectable rate.  However, we cannot exclude that it is a very slow and$/$or intermittent accretor.   

\subsection{Disk Properties} \label{dust}

\begin{figure}
\epsscale{1.1}
\plotone{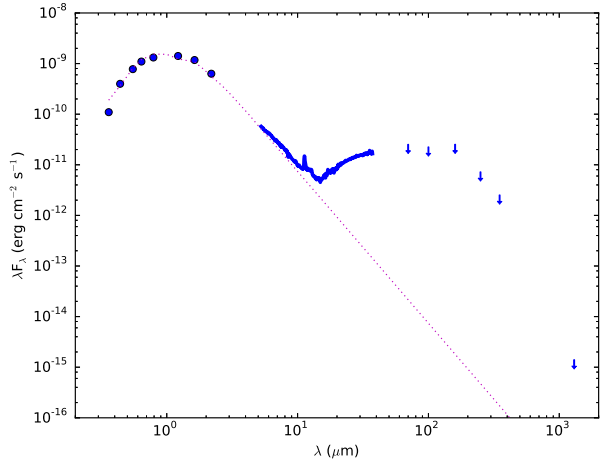}
\caption[]{
Spectral energy distribution of T~54 (blue).  
We show optical and infrared photomery (circles) and upper limits from {\it Herschel} PACS and SPIRE photometry and from ALMA (arrows).  We also show {\it Spitzer} IRS spectra (blue solid line).  All data is dereddened as described in Section~\ref{star} by fitting to a K5 photosphere (magenta dotted line).  Observational uncertainties are smaller than the sizes of the points used.  Mean uncertainties for the IRS data are $\sim$9$\%$ and are not shown for clarity. 
}
\label{figsed}
\end{figure}

Here we model the SED of T~54 to constrain its dust disk properties.  To compile the SED in Figure~\ref{figsed}, we take previously published data from the literature.  We use {\it Spitzer} IRS \citep{werner04, houck04} low-resolution spectra from the Cornell Atlas of Spitzer$/$IRS Sources \citep[CASSIS][]{lebouteiller11}\footnote{The Cornell Atlas of Spitzer/IRS Sources (CASSIS) is a product of the Infrared Science Center at Cornell University, supported by NASA and JPL.}.  We include optical photometry from \citet{rydgren80} 
and 2MASS JHK photometry from \citet{cutri03} as well as 
{\it Spitzer} IRAC and MIPS
data from \citet{luhman08b}.  In addition, we show far-IR {\it Herschel} PACS and SPIRE photometry from \citet{matra12}.  We present these data as upper limits since they 
include emission from a nearby FIR source, which we address further in the Appendix. 
We also show an ALMA upper limit at 1.3~mm from \citet{hardy15}.  We deredden the photometry and spectra with a visual extinction of 0.2 using the extinction law of \citet{mathis90} with an R$_{V}$ of 3.1.

The above data include both T~54~A and T~54~B. \citet{daemgen13} measured resolved JHKL photometry for each component.  We note that the colors based on resolved JHK photometry for T~54~A and T~54~B from \citet{daemgen13}
do not agree with the composite 2MASS colors for the system.  Given the difficulty of making such observations and the possibility for intrinsic stellar variability in one or both sources, we work with the composite data moving forward until there is more clarity on this issue.  We also use the stellar parameters of T~54~A in our modeling of the system since T~54~A dominates the composite emission \citep{lafreniere08,daemgen13}.   Therefore, T~54~A should dominate the heating of the disk.
We also note that there a polycyclic aromatic hydrocarbon (PAH) feature in T~54's IRS spectrum. We will return to this point in {\S}~\ref{pahs}.

\begin{figure}
\subfloat{
\includegraphics[width=\linewidth]{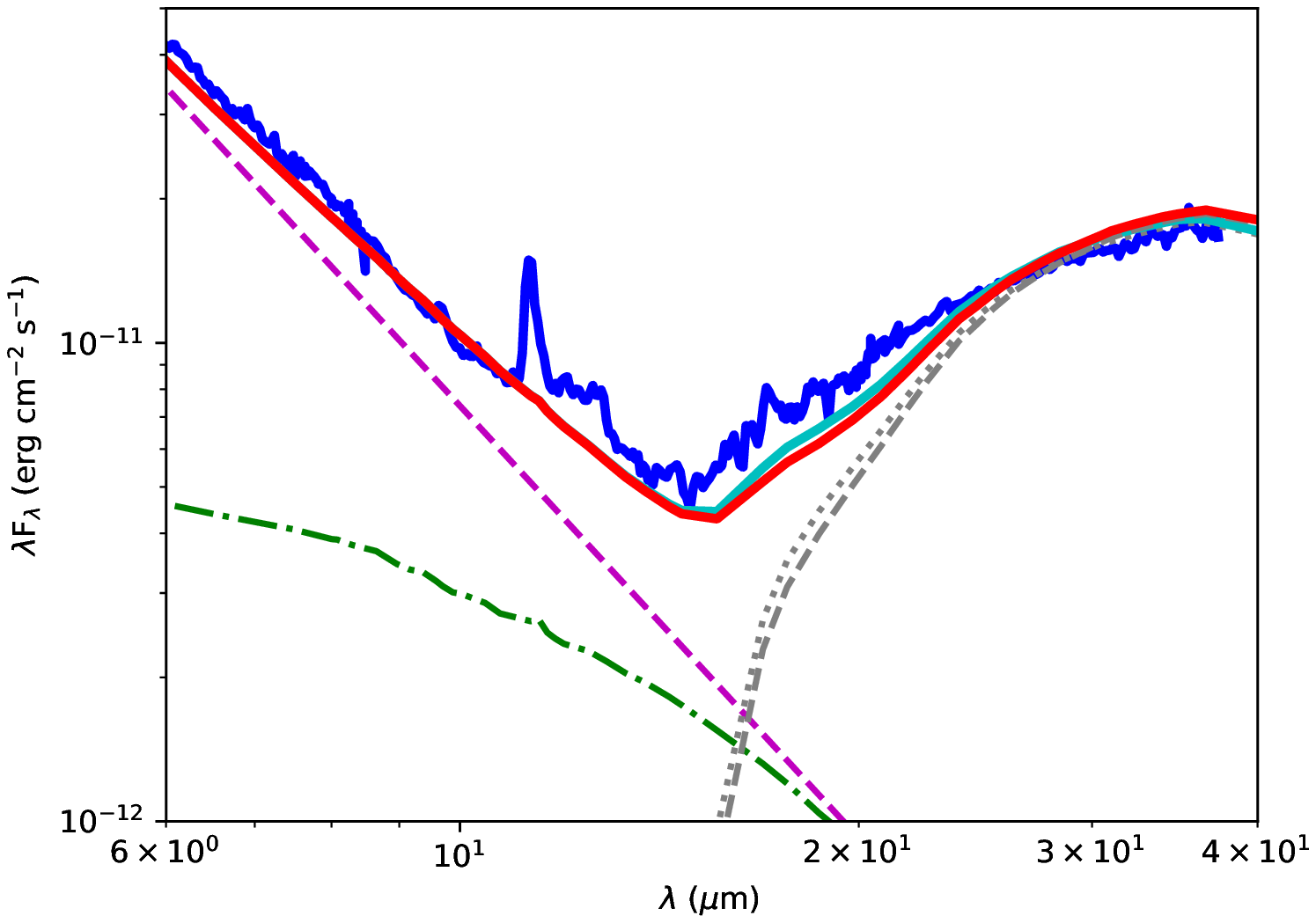}
}

\subfloat{
\includegraphics[width=\linewidth]{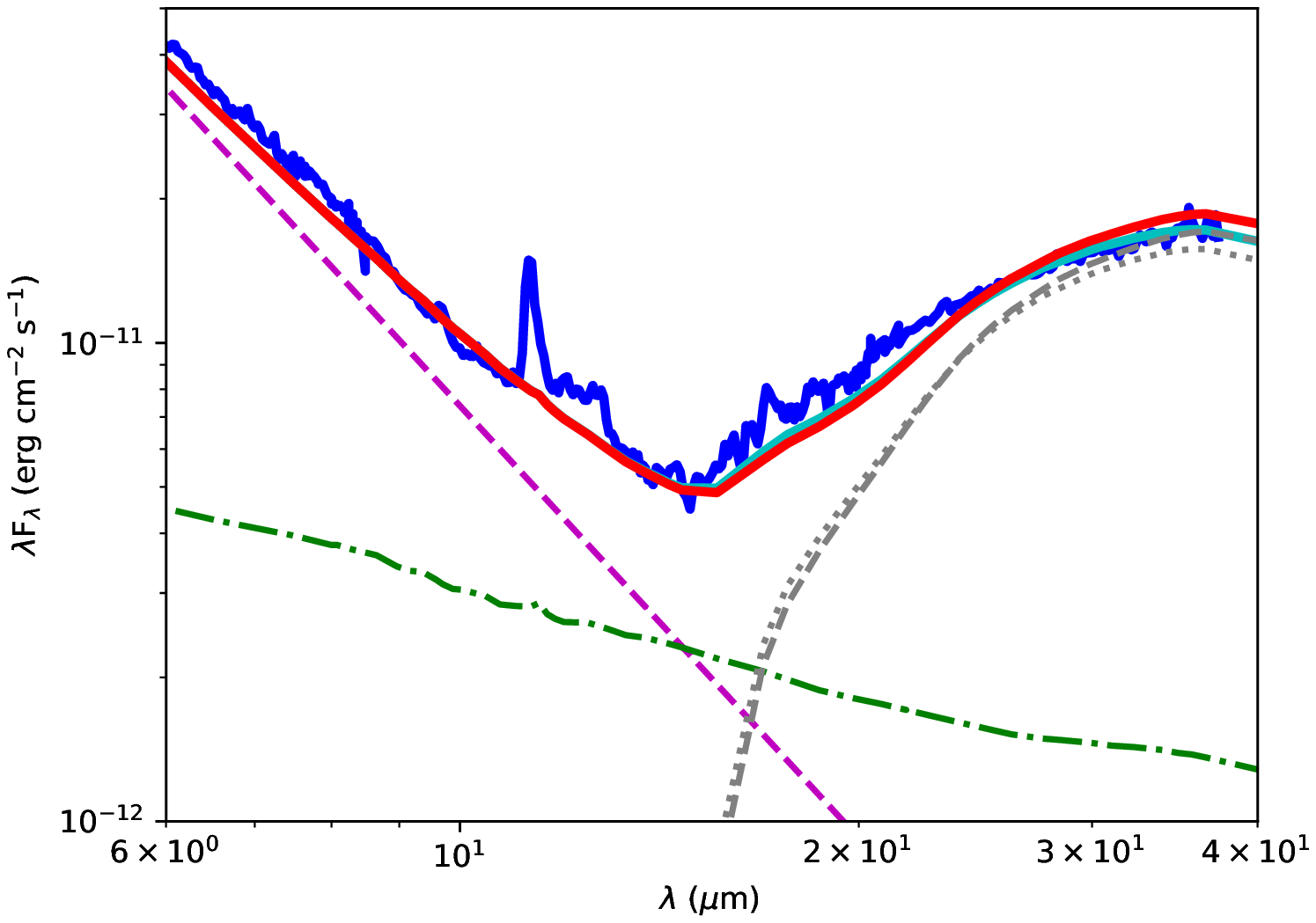}
}
\caption[]{
Top: IRS spectrum of T~54 (blue solid line) along with our best fitting models (solid lines). 
Each model consists of emission from a K5 photosphere (magenta dashed line), a circumprimary disk with an outer radius of 0.5 AU (green dashed dotted line), and a ``small'' circumbinary disk of ISM-sized grains (gray lines). We show the emission from circumbinary disks with radii of 60--100~AU (dotted) and 70--100~AU (dashed). 
The best fitting models (solid lines) include either the circumbinary disk with radii of 60--100~AU (cyan) or 70--100~AU (red). 
Bottom: Symbols and colors are the same as the top panel, except that we use the ``large'' circumprimary disk (green dashed dotted line) which has an outer radius of 13~AU.
}
\label{figsedmodel}
\end{figure}

\subsubsection{Modeling}

T~54 has excess emission above the stellar photosphere beginning at about 8~{\microns} which turns upwards at $\sim$15~{\microns} (Figure~\ref{figsed}).  
We were not able to reproduce this excess emission using models of optically thick disks \citep{dalessio06} because too much infrared emission was produced by those models.  Therefore, we used models of optically thin dust to reproduce the SED following \citet{calvet02} and \citet{espaillat11}.  Optically thin dust within $\sim$1 AU of the star has been used in the past to account for the NIR excess and 10~{\micron} silicate emission feature present in some transitional disks around TTS \citep[e.g.,][]{calvet05,espaillat07b}.  
Optically thin dust models have also been used to fit infrared excesses observed in {\it Spitzer} IRS spectra of debris disks \citep[e.g.,][]{matthews14, chen14, jang-condell15}.

In the optically thin limit, the temperature at a given disk radius $T_{0}$ can be solved following \citet{calvet02} using 
$\kappa_{P}(T_0)T_{0}^{4}R^{2} \sim \kappa_{P}(T_{*})T_{*}^{4}(z)R_{*}^{2}$ 
where $\kappa_{P}$ is the Planck mean opacity and $z$ is the height above the midplane. 
We adopt a dust composition of silicates and graphite, corresponding to that proposed by \citet{draine84} for the diffuse interstellar medium (ISM).  For silicates, the standard abundance (i.e., silicate dust-to-gas mass ratio or m$/$m$_{H2}$) is 0.004 with a dust sublimation temperature of 1400~K.  Graphites have an abundance of 0.0025 and a dust sublimation temperature of 1200~K.  Our model uses a dust grain size distribution following $a^{-p}$ where $a$ is the dust grain radius, which here ranges from 0.005~{\micron} to 0.25~{\micron} (i.e., ISM-sized), and $p$ is 3.5 \citep{mathis77}.  We assume spherical grains and construct the silicate and graphite dust opacities using Mie theory and optical constants from \citet{dorschner95} and \citet{draine84}, respectively.  
We assume the dust is evenly distributed (i.e., $\Sigma$ is constant with radius) since we do not have spatially resolved observations.  To fit the SED, we vary three parameters: the inner and outer radii of the disk (R$_{out}$, R$_{in}$) and $\tau$, the vertical optical depth evaluated at 10~{\micron}. 
We calculate the dust mass following 
M$_{disk}$=$\pi$ ($\Sigma$) (R$_{out}$$^{2}$-R$_{in}$$^{2}$).
As inputs to the model, we use the stellar parameters and distance for T~54~A noted in the previous section. 
  
The MIR excess short-wards of 15~{\micron} in Figure~\ref{figsed} indicates that some grains exist close to the star.  We assume a circumprimary disk is responsible for this emission.
To explain the MIR excess long-wards of 15~{\micron} in Figure~\ref{figsed}, we employ a circumbinary disk. 
In the following, we discuss our circumprimary and circumbinary disk SED modeling in more detail.
The best-fitting models are shown in Figure~\ref{figsedmodel} and parameters are listed in Table~2.  

\paragraph{Circumprimary Disk}

There is no evidence of a 10~{\micron} silicate emission feature in the IRS spectrum, indicating that no substantial amount of small grains are present close to the star.  We find that no more than 1.46$\times$10$^{-14}$ ${\msun}$ ($\sim$5$\times$10$^{-9}$ M$_{\Earth}$) of ISM-sized dust grains can exist within 0.5~AU or else a silicate feature would be seen.  Since there is no 10~{\micron} feature, the grains close to the star should be larger than 10~{\micron}.  However, with currently available data we cannot constrain if these grains are 10~{\micron}, 1~mm, or 1~cm in size or if there is a range of large grains present.   Therefore, we aim to set a lower limit to the  dust mass in the circumprimary disk and so model the circumprimary disk as composed of single-sized dust grains of 10~{\micron}.  If larger grains are present, the mass of the disk would be larger than reported below.  

We fix the inner radius to 0.1~AU to be consistent with the dust destruction radius of 10~{\micron}-sized grains around a K5 star and run models with outer radii (R$_{out}$) of 0.5~AU and 13~AU (Figure~\ref{figsedmodel}, top and bottom panels, respectively).  This is done to derive illustrative dust masses for a ``small'' and ``large'' size of the circumprimary disk.  Our choice for the largest radius to try comes from the expected truncation radius from binary disk clearing models \citep{artymowicz94, pichardo05}. Given the uncertainties in the binary separation reported by \citet{lafreniere08} and \citet{daemgen13} as well as those in the distance to Cha I \citep{whittet97}, the semimajor axis ($a$) of the binary can range from 34~AU -- 44~AU which would lead to truncation radii of the circumprimary disk in the range 10~AU -- 13~AU \citep[0.3$a$ for a binary with mass ratio 3:1;][]{artymowicz94}. We note that since T~54~A is significantly brighter than T~54~B we do not model a circumsecondary disk.  Although the existence of a circumsecondary disk is possible according to theory \citep{artymowicz94}, it would be too faint to detect against the emission of T~54~A, preventing the extraction of useful constraints.  

For each R$_{out}$, we vary $\tau$ in increments of 0.001 between 0.001 to 0.1 until we achieve a good fit to IRS spectrum.  The best-fitting $\tau$ for both the ``small'' and ``large'' circumprimary disk was 0.04.  The mass of dust in the best-fitting ``small'' circumprimary disk is $\sim$4${\times}$10$^{-12}$ ${\msun}$ and in the best-fitting ``large'' circumprimary disk it is $\sim$ 3${\times}$10$^{-9}$ ${\msun}$ ($\sim$10$^{-6}$--10$^{-3}$ M$_{\Earth}$  or 0.00001--0.08 lunar masses).  Again, this should be taken as a minimum mass since larger grains could be present that do not contribute to the SED.  Also, we do not claim this to be a unique model.  It should only be taken as illustrative of the minimum dust mass needed to reproduce the NIR excess in T~54.

\paragraph{Circumbinary Disk}

Here we use ISM-sized dust since only small grains are necessary to fit the IRS spectrum and we have no useful limits at longer wavelengths to constrain possible emission from larger dust grains.  Therefore, we do not have information to conclude whether or not there are large grains in the circumbinary disk.  We can only conclude that they are not necessary to reproduce the currently available SED.  Detections at longer wavelengths in the future can enable further exploration of the large dust grain content of the disk.  Therefore, our circumbinary disk mass measured here should be taken as a maximum mass of ISM-sized dust in the disk and a minimum mass of the total circumbinary disk mass.

A binary system with a circular orbit ($e$=0) can clear a hole in the disk of 1.8$a$ \citep{artymowicz94, pichardo05}. 
For the range of binary separations mentioned above, this would imply an inner radius of the circumbinary disk between $\sim$60--80~AU.
Accordingly, we vary the inner radius R$_{in}$ between 60 to 80 AU in increments of 5 AU and vary $\tau$ between 0.001 to 0.1 in increments of 0.001.  
We cannot well-constrain R$_{out}$ given that our data in the FIR and mm wavelengths are all quite high upper limits.  Therefore, we set R$_{out}$ to 100~AU and 200~AU. 
We combine the circumbinary disk model with the circumprimary disk model with outer radii of 0.5~AU (``small'') and 13~AU (``large'') 
and calculate the reduced $\chi^{2}$ for each model. 
The best fitting circumbinary disks have an R$_{in}$ of 60--70 AU and an outer radius of 100~AU (Figure~\ref{figsedmodel}).  In Table~2, we list the parameters for the best-fitting circumbinary disk models and denote whether they were paired with the ``small'' or ``large'' circumprimary disk. 
We find a mass range of $\sim$2--3${\times}$10$^{-8}$ ${\msun}$ ($\sim$8$\times$10$^{-3}$ M$_{\Earth}$ or 0.6 lunar masses).
Overall, models with inner radii greater than 70~AU or outer radii of 200~AU did not have enough  emission present at $\sim$20--30~{\micron} to reproduce the observed SED.

We do not claim our best-fitting models to the SED are unique fits.   However, we can fit the SED of T~54 with a circumbinary disk as well as circumstellar dust which has to be in large grains (10~{\micron} or greater) since no silicate feature is produced.
We cannot precisely constrain how the dust is distributed without spatially resolved observations, but our results are consistent with theoretical expectations of dynamical disk clearing by companions \citep{artymowicz94}.  

\begin{deluxetable}{l c c c c}
\tabletypesize{\scriptsize}
\tablewidth{0pt}
\tablecaption{Circumbinary Disk Model Properties}
\startdata
\hline
\hline
\multicolumn{5}{c}{Models with a Small Circumprimary Disk} \\
\hline
\colhead{R$_{in}$ (AU)} & \colhead{R$_{out}$ (AU)} & \colhead{$\tau$} & \colhead{M$_{dust}$ (M$_{\sun}$)} & \colhead{$\chi$$^{2}$}\\
\hline
60 & 100 & 0.017  & 2.2$\times$10$^{-8}$ & 2.42\\
65 & 100 & 0.021  & 2.5$\times$10$^{-8}$ & 2.79\\
70 & 100 & 0.025  & 2.6$\times$10$^{-8}$ & 3.06\\
\hline
\hline
\multicolumn{5}{c}{Models with a Large Circumprimary Disk} \\
\hline
\colhead{R$_{in}$ (AU)} & \colhead{R$_{out}$ (AU)} & \colhead{$\tau$} & \colhead{M$_{dust}$ (M$_{\sun}$)} & \colhead{$\chi$$^{2}$}\\
\hline
60 & 100 & 0.015  & 2.0$\times$10$^{-8}$ & 1.77\\
65 & 100 & 0.019  & 2.2$\times$10$^{-8}$ & 1.91\\
70 & 100 & 0.023  & 2.4$\times$10$^{-8}$ & 2.11
\enddata
\end{deluxetable}

\section{DISCUSSION} \label{sec:discuss}

\subsection{Gas in T~54}

\subsubsection{H$\alpha$} \label{sec:discusshalpha}

The cause of T~54's variable H$\alpha$ line profile is unclear.
Given that TTS are known to display strong flaring activity \citep[e.g.,][]{gahm90, guenther99, stelzer00},
one possibility is that T~54 was undergoing a flare during the observation where we detected an H$\alpha$ emission line.   
One particular flaring star has been well-studied: the K3 WTTS V410~Tau \citep[e.g.,][]{fernandez04, skelly10}.  \citet{fernandez04} found strong H$\alpha$ emission (EW$\sim$27~{\AA}) along with emission lines of H${\beta}$, H${\gamma}$, H${\delta}$, \ion{He}{2}, \ion{Fe}{2}, and \ion{He}{1} in an observation of V410~Tau and attributed these to a stellar flare. We note that in the spectrum where we detect H$\alpha$ in emission (2008 Feb 18) we did not detect H${\beta}$, H${\gamma}$, H${\delta}$, \ion{He}{2}, \ion{Fe}{2}, or \ion{He}{1}.  However, if one were to scale down the flux expected in most of these lines based on the observed H$\alpha$ emission they would be undetectable except for H${\delta}$, which \citet{fernandez04} observed to have a comparable flux to H$\alpha$. Therefore, we cannot exclude that T~54 was undergoing a weak flare at the time of our observations.  

Young stars also display evidence of smaller scale chromospheric activity such as microflares or prominences.
Studies of H$\alpha$ emission from WTTS find they have variable H$\alpha$ emission likely due to chromospheric activity \citep{scholz07}.  
In at least one observation of H$\alpha$ in V410~Tau, both emission and redshifted absorption were present which could only be explained with prominences \citep{skelly10}.  
\citet{fernandez04} found that in multiple quiescent spectra of V410~Tau only H$\alpha$ appears in emission and it has a variable line profile.  After calculating residuals, \citet{fernandez04} extracted a narrow component (EW$\sim$2--3~{\AA}) from the chromosphere along with a broad component that they connect to microflares.  
We note that in T~54 we do see a narrow component and faint traces of a broad component  which could possibly indicate the H$\alpha$ emission seen in T~54 is due to a microflare. 

However, a narrow H$\alpha$ profile with a broad component is also indicative of slow accretion \citep{espaillat08b}. In $\S$~\ref{sec:halpha} we measured an equivalent width (EW) at 10\% of the maximum flux of $\sim$200 km s$^{-1}$ for the self-subtracted H$\alpha$ profile residual of T~54. This value has been used as a cutoff for accretion for very low mass stars \citep{jayawardhana03}. 
\citet{hernandez14} reported the detection of several objects with
10\% EW($H_{\alpha}$)$<$200~km~s$^{-1}$ and IR excesses similar to that observed in CTTS.  They suggested that either these stars have stopped accreting, they are in a passive phase in which accretion was temporarily halted, or accretion is occurring below measurable levels.
While we do not see evidence of substantial, continuous accretion in T~54, a circumprimary and circumbinary disk are possibly present in the system, and so, we cannot rule out weak episodic accretion occurring in T~54.  Sporadic accretion events can continue into the debris disk regime. For example,   falling evaporating bodies (FEBs) have been observed in debris disks.  These events are thought to be infalling comets seen seen via variable redshifted absorption features as gas is produced that transits in front of the central star \citep[e.g.,][]{vidal-madjar94, welsh13, kiefer14, eiroa16}.   To explore episodic accretion in T~54 further, high-resolution spectra with much finer time sampling than in this work would be needed.   

\subsubsection{Polycyclic Aromatic Hydrocarbons} \label{pahs}

There is a relatively strong 11.3~{\micron} feature in T~54's IRS spectrum (Figures~\ref{figsed} \&~\ref{figsedmodel}). 
Polycyclic aromatic hydrocarbons (PAHs) are known to display a $\sim$11~{\micron} feature which arises from the C--H bending modes in vibrational models of PAHs \citep{draine07}.
Alternatively, there is a 11.3~{\micron} SiC feature is seen in carbon stars.  However, this SiC feature is typically much broader than what is seen in T~54 \citep[e.g.,][]{papoular98,fujiyoshi15}.  Therefore, we attribute the 11.3~{\micron} feature in T~54's IRS spectrum to PAHs.

Interestingly, we do not see the PAH features at $\sim$6~{\micron} and $\sim$8~{\micron}, which usually accompany the $\sim$11~{\micron} feature.
In the case of ionized PAHs, the $\sim$6~{\micron} and $\sim$8~{\micron} features are stronger than the $\sim$11~{\micron} feature; for neutral PAHS the $\sim$11~{\micron} feature is stronger than the others \citep{draine07}.  One can speculate that T~54 has neutral PAHs and that the $\sim$6~{\micron} and $\sim$8~{\micron} features are hidden by the photophere and inner disk emission.

PAHs are usually seen in early-type stars; to date, no T Tauri stars with spectral types later than G8 show the PAH emission features in their MIR spectra \citep{tielens08}.  We note that \citet{furlan06} reported the PAH feature at $\sim$11~{\micron} in UX~Tau~A.
\citet{espaillat10} later classified UX Tau~A as a G8.0$\pm$2.0
star.  Previously, \citet{herbig77} and \citet{cohen79} both reported a spectral
type of K2,  \citet{rydgren76} reported G5, and \citet{hartigan94} reported
K2--K5.
To the best of our knowledge, T~54 is the first detection of a 11~{\micron} PAH feature around a late-K-type star. This is suggestive that PAHs are present around K-type and M-type TTS as well, but perhaps hidden by the continuum and 10~{\micron} silicate emission feature, or that they are tied to the amount of UV radiation penetrating the disk atmosphere, which will be greater as the dust settles and$/$or grows. 

It is unclear where in the disk the PAHs are originating from, but they have been connected to gas flows from the disk onto the star in some cases.
\citet{maaskant14} can fit the PAH spectra in their sample of four transitional disks surrounding accreting stars with a combination of ionized PAHs located in gas flows traversing the low-density disk cavity in the inner disk and neutral PAHs originating in the outer disk.
In at least one transitional disk \citep[IRS~48;][]{brown12}, spatially resolved MIR images detect PAHs within the disk cavity \citep{geers07}.  
Spatially resolved data of the PAHs in T~54 would reveal where the PAHs are located in the system.

\subsubsection {[\ion{Ne}{2}] and [\ion{Ne}{3}]}

\citet{espaillat13} reported both [\ion{Ne}{2}] and [\ion{Ne}{3}] line emission from T~54. 
Theoretical predictions show that MIR Ne line emission originates from within the inner disk given that X-ray photons cannot heat the gas to high enough temperatures (T$\sim$4000~K) to create Ne line emission tens of AU  from the star \citep[e.g.,][]{glassgold07}.  Using X-ray luminosities ranging from 2$\times$10$^{29}$ erg s$^{-1}$ to 2$\times$10$^{31}$ erg s$^{-1}$, the farthest disk radius producing Ne line emission is about 25 AU \citep{meijerink08}. T~54 has an X-ray luminosity of $\sim$8$\times$10$^{30}$ erg s$^{-1}$ \citep{ingleby11b,feigelson93}, within the range probed by \citet{meijerink08}.
This suggests that T~54 has gas in the inner region, within at most 25 AU.  The existence of a circumprimary dust disk is consistent with this.  Alternatively, Ne emission has been connected to jets \citep{gudel10}, but given that jets are tied to accretion \citep{rigliaco13}, it is unlikely the very slowly or non-accreting T~54 has a powerful enough jet to lead to Ne emission.

We can calculate the minimum gas mass necessary to create the observed Ne line emission following \citet{glassgold07}.   
We write the luminosity of the line as:
\begin{equation}
L=\int{\rm d}VX_{II}n_{\rm Ne}P_uA_{ul}h\nu_{ul}
\end{equation}
where $P_u$ is the population of the upper level, $A_{ul}$ is the Einstein A Co-efficient and $h\nu_{ul}$ is the energy of the transition. $X_{II}$ is the fraction of Ne atoms in the Ne$^+$ state and $n_{\rm Ne}$ is the number density of Ne atoms in a volume $V$. The population level of the upper level is given by:
\begin{equation}
P_u=\frac{1}{2C_{ul}e^{(1122.8/T)}+1}
\end{equation}
where $C_{ul}$ = 1 + n$_{cr}$($ul$)$/$n$_{e}$ and accounts for effects of the critical density, n$_{cr}$.  This means $C_{ul}\rightarrow 1$ when $n<n_{cr}$ and $P_u\rightarrow 0$ when $n>n_{cr}$. For a constant temperature medium we can then approximately write:
\begin{equation}
L\approx\frac{A_{ul}h\nu_{ul}}{2e^{(1122.8/T)}+1}\int n_{Ne+}(<n_{cr}){\rm d}V
\end{equation}
Thus, given a measured luminosity we can use the above expression to give an estimate of the mass in Ne as a function of temperature. In the case $T\gg1122.8$ then the result is independent of temperature. Or,
\begin{equation}
L\sim\frac{A_{ul}h\nu_{ul}}{3}\int n_{Ne+}(<n_{cr}){\rm d}V
\end{equation}
Taking the value of $A_{ul}h\nu_{ul}=1.332\times10^{-15}$ erg s$^{-1}$ from \citet{glassgold07} then we can compute minimum masses. Using a Ne$/$H number ratio of $X_{\rm Ne}=1.2\times10^{-4}$ for a solar-like mixture \citep{ercolano10}, then the minimum gas mass is approximately:
\begin{equation}
M_{\rm min}\approx \frac{m_H}{X_{II}X_{\rm Ne}}\int n_{Ne+}(<n_{cr}){\rm d}V
\end{equation}
thus,
\begin{equation}
M_{\rm min}\sim 7\times10^{-8} {\,\rm M_{J}\,}X_{II}^{-1}\left(\frac{L_{\rm [NeII]}}{10^{-6} L_{\odot}}\right).
\end{equation}

With both \ion{Ne}{2} and \ion{Ne}{3} lines one can estimate the level populations. From (2.11) in \citet{glassgold07}, assuming the critical densities are similar then the electron fraction is approximately $X_e\sim N_{\rm Ne2+}/N_{\rm Ne+}$. Or the electron fraction is
\begin{equation}
X_e\sim\left(\frac{L_{\rm [NeIII]}}{L_{\rm [NeII]}}\right)\left(\frac{A_{\rm [NeII]}h\nu_{\rm [NeII]}}{A_{\rm [NeIII]}h\nu_{\rm [NeIII]}}\right)\sim 0.57\left(\frac{L_{\rm [NeIII]}}{L_{\rm [NeII]}}\right)
\end{equation}
Now in the case $X_e\ll1$ then $X_{\rm II}\sim \frac{\zeta(Ne)}{\zeta} X_e$, with $\zeta(Ne)$ the ionization rate of Ne and $\zeta$ the ionization rate of the gas. In the case that $X_e$ is not $\ll 1$ then we can assume $X_{II}\sim 1$ and $X_{III}$ (the fraction of Ne atoms in the Ne$^2+$ state) is $\sim X_e$. Taking the values from \citet{espaillat13} for the [\ion{Ne}{2}] and [\ion{Ne}{3}] line luminosities we find $X_e\sim 1/6$. Since the ionization rate of Ne is much larger than Hydrogen (of order 40 for a 1~KeV X-ray spectrum), but the recombination rates are similar, then one would expect $X_{II}$ is of order unity. For simplicity, we set $X_{II}\sim 1$ which gives us $X_{III}\sim 1/6$. 
This makes the [\ion{Ne}{3}] line more constraining for mass. 
We find that for solar mixtures the minimum mass of gas is 
$\sim$10$^{-9}$ $M_{\sun}$ ($\sim$10$^{-6}$ M$_{Jup}$ or $\sim$3$\times$10$^{-4}$ M$_{\Earth}$), 
or $\sim$10$^{-12}$ $M_{\sun}$ ($\sim$10$^{-9}$ M$_{Jup}$ or $\sim$3$\times$10$^{-7}$ M$_{\Earth}$) in Ne gas.

\subsubsection{[\ion{O}{1}]} \label{OI}

T~54 also has an [\ion{O}{1}] 63~{\micron} line detection \citep[see Appendix;][]{keane14,riviere-marichalar16}. 
The GASPS survey \citep{howard13} found that the [\ion{O}{1}] 63~{\micron} line is the strongest FIR line seen in disks.  
Theoretical simulations predict that in protoplanetary disks this line is dominated by emission from outside of 30~AU, but in optically thin disks this emission can originate closer to the star \citep{kamp10}. 
Therefore, the origin of FIR [\ion{O}{1}] emission in T~54 may be a circumprimary disk, circumbinary disk, or both.  Alternatively, \citet{howard13} find that sources with jets have [\ion{O}{1}] 63~{\micron} line fluxes that are an order of magnitude higher than sources without known jets.  However, as noted above in the case of Ne emission, it is unlikely there is a strong enough jet in T~54 to lead to the observed [\ion{O}{1}] emission.

\citet{howard13} report that the [\ion{O}{1}] 63~{\micron} line is fainter in transitional disks than in full disks and do not detect this line in Class III disks. 
For T~54, \citet{keane14} report a line flux of 4.5$\pm$0.2$\times$ 10$^{-17}$ W m$^{-2}$ and a 63 {\micron} continuum flux of 0.60$\pm$0.02 Jy. \citet{riviere-marichalar16} measure a line flux of 2.4$\pm$0.1$\times$10$^{-17}$ W m$^{-2}$ with a 63 {\micron} continuum flux of 0.33$\pm$0.09 Jy. 
Using this range of line flux estimates, T~54's [\ion{O}{1}] emission is at the brighter end of the [\ion{O}{1}] emission emission seen in transitional disks.  
We speculate this is because the disk of T~54 is optically thin; the low amount of dust allows high-energy radiation fields to penetrate the disk more deeply and heat more of the disk, and so we see more of the gas that is present relative to optically thick disks. 
Based on \citet{woitke10}, for the observed  [\ion{O}{1}] 63~{\micron} line fluxes reported, T~54 could have 10$^{-6}$ --10$^{-3}$ {\msun}  ($\sim$0.3--300 M$_{\Earth}$)
of gas for temperatures of 32--70~K.  We use this as a rough estimate only given that the calculations of \citet{woitke10} are for primordial disks with higher densities and less penetration of high-energy radiation.

\subsubsection{CO}

\citet{hardy15} observed the $^{12}$CO~(2--1) line with ALMA at 1.3~mm and did not detect the line nor the continuum.  They had a spectral resolution of 976.56~kHz, 
equivalent to about 1.3 km s$^{-1}$ at 230 GHz, and a sensitivity of 90 mJy (3$\sigma$ per channel).  
Using the ALMA non-detection, we can estimate an upper limit to the gas mass of the disk.  

We assume the $^{12}$CO~(2--1) line is optically thin and that the disk is spatially unresolved.  We write the luminosity of the line as
\begin{equation}
L=\int{\rm d}V\left(\frac{n_u}{n_{CO}}\right)n_{\rm CO}A_{ul}h\nu_{ul}
\end{equation}
where $V$ is the volume of the disk emitting region using radii spanning 60--200~AU.
Here $n_u$ is the population of the upper level; assuming a standard CO freeze-out temperature of 20~K, we adopt an excitation temperature of 30~K.
$n_{CO}$ is the number density of CO assuming that all the C is in CO and a CO abundance relative to H$_{2}$ of 3.63$\times$10$^{-4}$.
$A_{ul}$ is the Einstein coefficient
and $h\nu_{ul}$ is the transition energy.
We estimate an upper limit to the gas mass of 5$\times$10$^{-3}$ {\msun} ($\sim$200 M$_{\Earth}$).

\subsubsection{Gas-to-dust mass ratio}

From the above, the picture that emerges of T~54 is that of a system that has some gas in both its  circumprimary and circumbinary disk.  
The upper limit to the total gas mass from the CO observations (5$\times$10$^{-3}$ {\msun}) is consistent with the rough gas mass estimate inferred from the [\ion{O}{1}] line flux (10$^{-6}$--10$^{-3}$ {\msun}) and the minimum mass needed to create the observed Ne emission ($\sim$10$^{-9}$ {\msun}).  
Using our derived range of minimum dust masses in T~54 ($\sim$2--3$\times$10$^{-8}$ {\msun}) and our minimum gas mass from the Ne line, this suggests a gas-to-dust ratio of about 0.1.  This should be taken as illustrative, but not precise given that if there is a substantial amount of unseen large grains in the disk the gas-to-dust mass ratio would be smaller and if there is much more gas than implied from the Ne observations the gas-to-dust mass ratio would be larger. 

Lower gas-to-dust mass ratios in protoplanetary disks have been proposed previously. 
Using millimeter wavelength CO lines, \citet{williams14} and \citet{ansdell16} measure gas-to-dust ratios that are an order of magnitude less than the typically assumed 100.  One particular case is the transitional disk of TW~Hya where \citet{williams14} find a gas-to-dust ratio of 4 and \citet{thi10} inferred a gas-to-dust ratio of $\sim$30 using FIR {\it Herschel} PACS observations of the [\ion{O}{1}] line.  However, \citet{bergin13} derive a gas-to-dust ratio for TW~Hya of $\sim$100 using {\it Herschel} HIFI observations of the HD line, which should trace the bulk of gas in the disk \citep{mcclure16}. 
Interestingly, using the [\ion{C}{1}] millimeter line, \citet{kama16} found that gas phase carbon is depleted by a factor of about 100 in TW~Hya. This is suggestive that the CO abundance is lower in the outer disk of TW~Hya, which may account for the lower gas-to-dust ratio measured using CO.  
Clearly, more work needs to be done to decipher the gas-to-dust mass ratios of disks, which is complicated by the fact that it is difficult to distinguish between low gas-to-dust ratios and low CO$/$H$_{2}$ abundance \citep{miotello17}.  Regardless, the gas-to-dust ratio inferred here for T~54, using the highly sensitive Ne line instead of only CO, is much lower than seen in most protoplanetary disks.

\subsection{Dust in T~54}

Given that we can reproduce the SED of T~54 with circumprimary and circumbinary disk sizes that are roughly consistent with binary disk-clearing theory \citep{artymowicz94}, it is reasonable to conclude that dynamical clearing has occurred in this system.  The disk gap carved by the binary can be crossed by accretion streams and form small disks around the stars \citep{artymowicz96} and it is expected that these streams are too faint to be detected \citep{shi16}.  We speculate that the grains in the circumprimary disk are large due to dust growth and dust radial dust drift, as has been seen in other systems \citep[e.g.,][]{andrews15}. 

T~54's dust disk does not resemble what is seen in protoplanetary disks.
T~54 has a low disk fractional luminosity (i.e., the total flux from  the disk compared to flux from star) of $\sim$0.01 according to our modeling, similar to what is seen in debris disks \citep[e.g., $<$0.01][]{wyatt15}.  
Comparing the fractional excess of T~54 at 12~{\micron}, 22~{\micron}, and  70~{\micron} (F$_{tot}$$/$F$_{*}$=2, 14, $<$1000 respectively) to Figure 1 of \citet{wyatt15} we see that T~54 most resembles $\beta$ Pic.  T~54 lies well within the region proposed by \citet{wyatt15} to encompass debris disks (F$_{tot}$$/$F$_{*}$$<$ 3 at 12~{\micron} and F$_{tot}$$/$F$_{*}$$<$ 2000 at 70~{\micron}). We note that \citet{wyatt15} did this analysis of fractional excess only for A stars.  However, there appears to be no dependence between disk fractional luminosity and spectral type from late B to M stars on the main-sequence \citep{matthews14}.  Therefore, our comparison of T~54 to \citet{wyatt15}'s proposed classification scheme is not unreasonable.

\subsection{Evolutionary Stage}

T~54 is likely in a transient phase between the protoplanetary and debris disk stages.
Overall, there is some that evidence the dust and gas components of disks dissipate roughly concurrently based on a correlation found between emission of CO gas and dust continuum from CTTS, Herbig Ae and debris disks \citep{pericaud17}.  However, there are some disks that lie above the trend, suggesting a transient phase when the dust has evolved faster than the gas \citep[e.g.,][]{pericaud17}.  

The best-studied disk to date in this transient phase is 
the 5~Myr old HD~141569 triple system, whose primary is a Herbig Ae$/$Be B9.5 star \citep{weinberger99, fairlamb15}.  
HD~141569 has a fractional
disk luminosity of 0.0084 \citep{clampin03}, which is intermediate between old, evolved protoplanetary disks and young, luminous debris disks \citep{wyatt15}.
HD~141569 has evidence of small grains from 23--70~AU observed in the L-band \citep{mawet17} and millimeter observations indicate the presence of large dust out to about 56~AU \citep{white16}.  There are additional rings of dust in the outer disk at 100s of AU \citep{augereau99, weinberger99, konishi16}.
\citet{thi14} measure a dust mass of 2$\times$10$^{-7}$ ${\msun}$  ($\sim$0.1 M$_{\Earth}$), including up to 1~mm grains.
H$\alpha$ emission is detected within 0.12~AU of the primary,
but it is unclear if this originates from magnetospheric accretion \citep{mendigutia17}.
HD~141569 displays PAH features and detections of the [\ion{O}{1}] lines at 63~{\micron} and 145~{\micron}, and also the \ion{C}{2} line at 157~{\micron} \citep{thi14}.
There is hot gas in the inner disk ($<$50~AU) detected via ro-vibrational CO lines in the NIR \citep{brittain02, goto06, salyk11b}.  
There is also evidence of $^{12}$CO (3--2) gas at 30 -- 210~AU in the disk using ALMA \citep{white16}. 
Observations have constrained the total gas mass to 13 -- 200 M$_{\earth}$, mostly originating from the outer disk according to CO measurements \citep{thi14, flaherty16}.

T~54's disk fractional luminosity is similar to that of HD~141569.  
While one may conjecture that T~54 is in a similar evolutionary state based on its dust properties alone, it is difficult to draw conclusions since T~54 is not as well-studied in either its dust or gas distributions.
Interestingly, T~54 is only $\sim$2 Myr old, making this object the youngest debris disk that still harbors gas.
Given that disk evolution is slower around low-mass stars when compared to high-mass stars regardless of age \citep[e.g.,][]{hernandez05,ribas15},
it is likely that we are able to catch T~54 in an incipient debris disk phase at such a young age since its binary nature speeds up disk evolution.
In some cases, multiplicity has been found to play a role in disk evolution, namely by increasing disk dispersal and reducing disk lifetimes
\citep[e.g.,][]{daemgen15,kraus12,cieza09}.
In addition, debris disks are common around multiple stars \citep{monin07, matthews14} and the
fractional luminosity of dust in multiple systems is lower than in single systems, pointing to more efficient clearing in multiple systems \citep{rodriguez12,rodriguez13,matthews14}.

T~54 is one of the latest spectral type primary stars with an incipient debris disk  containing gas reported to date. TWA~34, an M5 star located in the 10~Myr old TW Hya Association, may be a similar object.  From millimeter detections of the dust continuum and CO gas, a relatively low disk mass of $\sim$3$\times$10$^{-2}$ M$_{\Earth}$ and an inferred gas-to-dust ratio of 10 have been measured for this object \citep{rodriguez15}.
Debris disks with gas are typically reported around earlier type stars.  For example, \citet{lieman-sifry16} find in the Scorpius-Centaurus OB association that in a sample of 23 debris disks, there are three with strong CO detections, all around A stars.

T~54 presents a unique opportunity to study an evolved disk in a young star-forming region.
The circumprimary disk in the T~54 system also offers an interesting connection to recently discovered multiple systems with tightly packed rocky planets close to the primary star \citep[Kepler~444; e.g.,][]{dupuy16}.   Kepler~444~A is a K0 main-sequence star with a pair of M dwarfs $\sim$66~AU away; five planets are located within 0.1~AU of the primary star. Since the binary companions should have truncated the protoplanetary disk at about 2~AU, the initial inner disk must have been quite massive in order to form multiple planets \citep{kratter10, dupuy16}. Some have proposed that Kepler~444 began its evolution with a massive inner disk in a triple system similar to that observed by \citet{tobin16}.  Over time the inner disk dissipated via accretion onto the star, photoevaporation, and planet formation.  T~54 may be what a system like Kepler~444 looks like after the end of its protoplanetary disk lifetime.

\section{SUMMARY \& CONCLUSIONS}

Here we presented new data and analysis of T~54, a $\sim$40~AU separation binary in Cha I.  Using high-resolution optical spectra taken with the MIKE instrument at {\it Magellan}, we measured a new spectral type of K5$\pm$1 for T~54~A, which is later than previous works reporting this object as a late-G or early-K star.  Accordingly, we also derived a younger age of $\sim$2 Myr for this system.  

We studied archival HST ACS FUV data and found no evidence for H$_{2}$ emission close to the star.  We also presented new {\it Magellan} FIRE NIR spectrographic observations of Br$_{\gamma}$, which was not detected, and of \ion{He}{1}, which lacked the high velocity red-shifted absorption seen in very slow accretors.  In one of our four epochs of MIKE spectra, we detected the H${\alpha}$ line in emission, suggestive that we caught a small burst of accretion.  

{\it Spitzer} spectra of T~54 indicate a rise in the MIR emission from nearly photospheric levels at shorter wavelengths to a weak excess above the stellar photosphere at longer wavelengths, an indication that some dust remains in the system.  
We modeled the SED of T~54 and found it could be reproduced with a circumbinary disk and a circumprimary disk in the system, with a combined dust mass of $\sim$10$^{-8}$ $\msun$ ($\sim$3$\times$10$^{-3}$ M$_{\Earth}$).  Both disks have radii consistent with theoretical predictions of dynamical clearing by stellar-mass companions \citep{artymowicz94}. 

There is previously published evidence of gas remaining in the disk indicated by [\ion{Ne}{2}],  [\ion{Ne}{3}], and [\ion{O}{1}] line detections.  The detection of Ne lines is particularly interesting since these lines should originate from within the inner few AU of disks.  Using the Ne lines we derive a minimum mass of gas in the disk of $\sim$10$^{-9}$ $\msun$ ($\sim$3$\times$10$^{-4}$ M$_{\Earth}$).  This suggests that ongoing accretion is occurring at such a slow rate that it is not detectable and that we caught H$\alpha$ in emission when the accretion was barely high enough to detect.  

Despite the evidence for some gas in the disk and its young age, T~54's disk fractional luminosity and low dust mass is consistent with what is seen in debris disks.  
We conclude that T~54 has ended its primordial disk lifetime and is crossing over the bridge to the debris disks stage.  The fact that this is a binary system hastens its evolution, allowing us a rare glimpse of this bridge in a young star-forming region.  Future observations of this object will help us understand this short-lived stage. 

 \acknowledgments{
We thank the referee for comments that helped to improve the paper. 
We thank S. Daemgen and L. Ingleby for sharing their published data. We appreciate helpful discussions with M. Hughes, K. Luhman, and J. Muzerolle.
This material is based upon work supported by the National Science Foundation under Grant Number AST-1455042 and the Sloan Foundation.
}

\appendix

\begin{figure*}
\centering
\includegraphics[width=0.36\hsize]{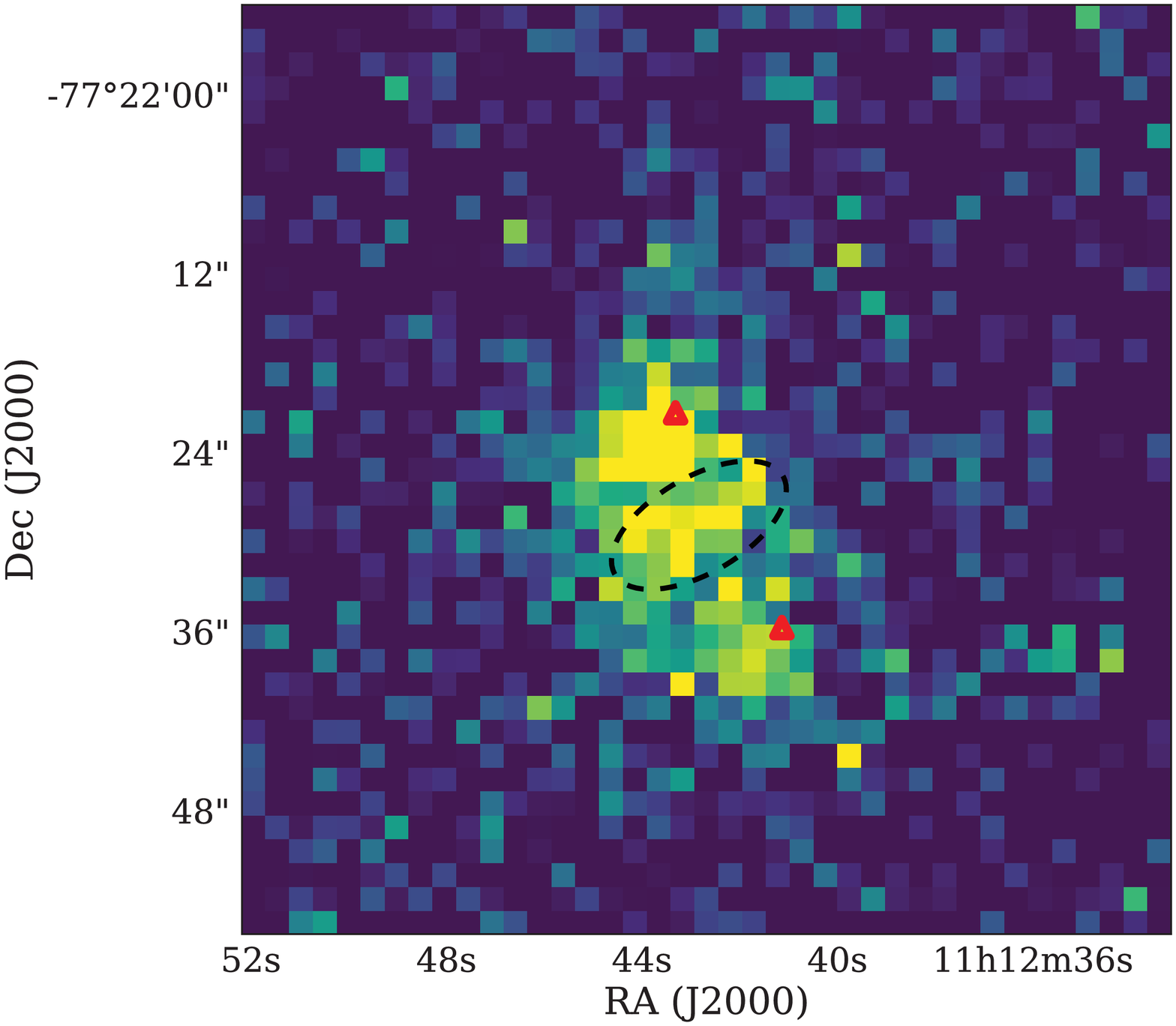} 
\includegraphics[width=0.29\hsize]{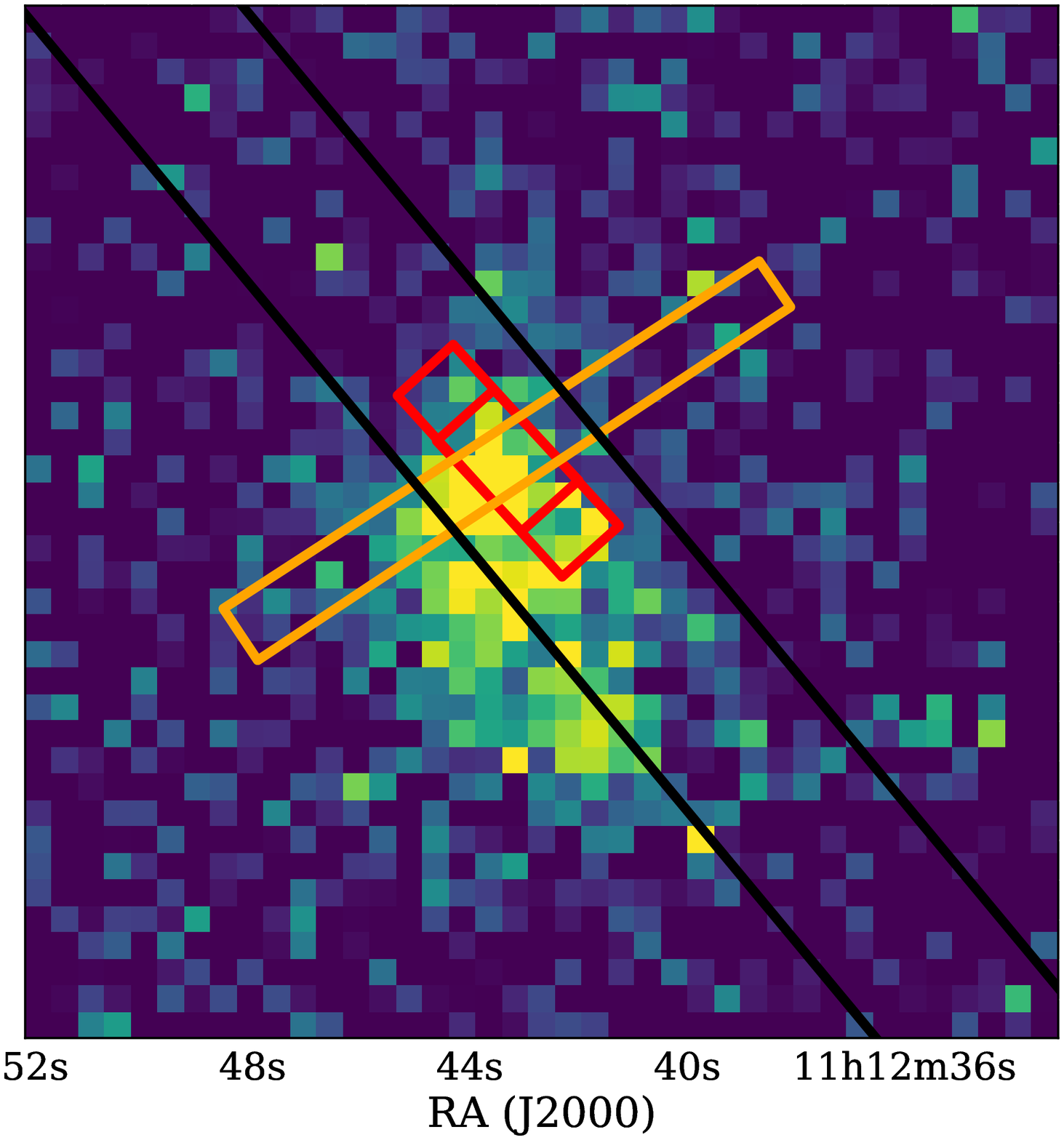} 
\includegraphics[width=0.29\hsize]{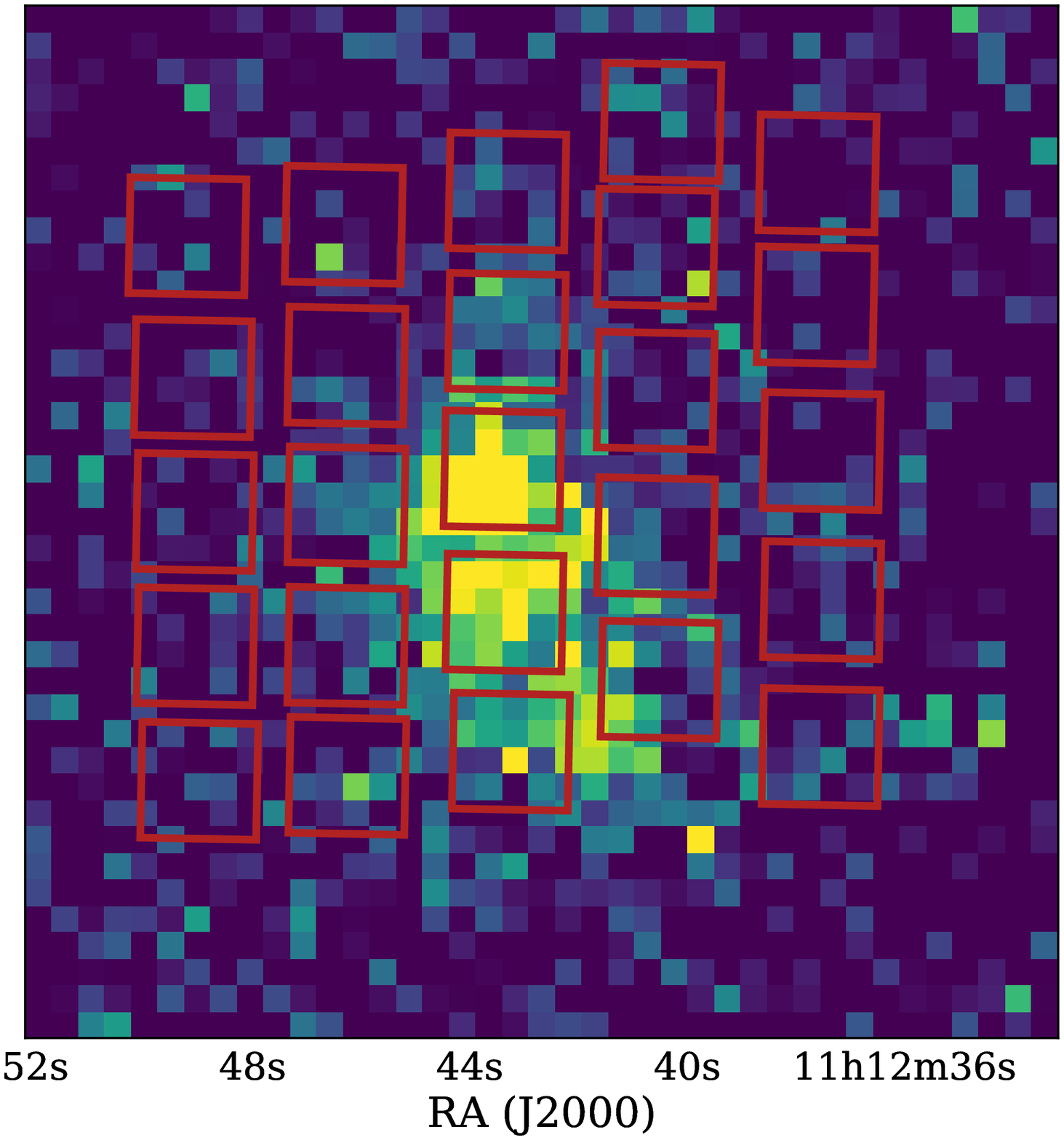}
\caption{Left: PACS 70~{\micron} image from the HSA archive with T~54 and a nearby 2MASS source labeled with red triangles (top and lower, respectively).  The FIR source identified by \citet{matra12} is denoted with a black ellipse.
Center: 
{\it Spitzer} IRS spectral footprint overlaid on the PACS 70~{\micron} image.  We show the Short-Low (yellow) and Long-Low (black) slits as well as the two nods of the Short-High observations (red).
Right:
{\it Herschel} PACS spectral footprint (red boxes) corresponding to the [\ion{O}{1}] observations discussed in Section~\ref{OI} overlaid on the PACS 70~{\micron} image.  Note that T~54 is in the central spaxel, which was used to extract the spectrum, and the unidentified FIR source is not covered by the central spaxel.}
\label{overlays}
\end{figure*}

T~54 has a nearby FIR source that contaminates the observed FIR emission \citep[left, Figure~\ref{overlays};][]{matra12}.  Here we show that this unidentified FIR source should have no significant contribution to most of the data that we study in this paper. The unidentified FIR source is 6$^\prime$$^\prime$ away from T~54.  Assuming it is in Cha I (at 160 pc), it is located 1040~AU away from T~54 at a position angle of 196$^{\circ}$. To the best of our knowledge, this FIR source does not overlap with previously reported objects.  

\citet{matra12} find that the unidentified FIR source is not detected in JHK-band photometry, IRAC photometry, or MIPS data taken at 24~{\micron}.  However, it is detected in PACS 70~{\micron} and 100~{\micron} data as well as SPIRE photometry taken at 160~{\micron}, 250~{\micron}, and 350~{\micron}.  The flux of this unidentified FIR source is comparable to or greater than T~54 at 100~{\micron} and beyond \citep{matra12}.  Therefore, in Figure~\ref{figsedmodel} we show the fluxes reported by \citet{matra12} as upper limits since these data include emission from both the T~54 system and the nearby, unidentified FIR source.

It is unlikely that the unidentified FIR source is contributing to the {\it Spitzer} IRS spectra of T~54.  As reported by \citet{matra12}, the unidentified FIR source is not detected in NIR nor MIR IRAC or MIPS images. To explore this further, we overlaid the IRS slit positions on a PACS 70~{\micron} image of T~54 (center panel, Figure~\ref{overlays}).  There are low-resolution and high-resolution IRS spectral observations of T~54 available in CASSIS.  In the low-resolution observations (2005-04-24 10:57:08), the Short-Low module covered only T~54 while the Long-Low module included the position of the unidentified FIR source.   In the high-resolution IRS observations  (2008-10-12 02:08:59), the Short-High module covers the position of the unidentified FIR source in only one of the two nods.  In the optimally extracted data of each nod from CASSIS \citep{lebouteiller10, higdon04}, the continuum in both nods agrees, supporting that the unidentified FIR source did not contribute to the IRS data.  
{\it Spitzer} has a blind pointing accuracy of 0.5$^\prime$$^\prime$ \footnote{https:$/$$/$irsa.ipac.caltech.edu$/$data$/$SPITZER$/$docs$/$spitzermission$/$missionoverview$/$spitzertelescopehandbook$/$Spitzer$\_$Telescope$\_$Handbook.pdf} so it is unlikely that the {\it Spitzer} observations were mispointed such that both nods included the unidentified FIR source, keeping in mind that the object is 6$^\prime$$^\prime$ away.
We note that the continuum in the low-resolution and high-resolution spectra agree. In addition, the Neon line emission detected in the high-resolution spectra is similar between both nods.  Therefore, we conclude that the MIR excess and Neon line emission are from the T~54 system and that there is no substantiative contribution from the unidentified FIR source.

Since the unidentified FIR source is detected at 70~{\micron} and beyond, we explored if it could be contributing to the [\ion{O}{1}] 63~{\micron} line. In Figure~\ref{overlays} (right), we overlay the {\it Herschel} PACS spectral footprint on the 70~{\micron} PACS image. The central spaxel, from which the PACS spectrum is extracted, does not include the unidentified FIR source.  However, the pointing accuracy of {\it Herschel} does allow for the possibility that the unidentified FIR source was included in the central spaxel.
The PACS image was observed on OD 219 and the spectrum was taken on OD 933, when the pointing accuracies were reported to be $\sim$2$^\prime$$^\prime$ and $\sim$1$^\prime$$^\prime$, respectively. \footnote{Based on numbers quoted in http:$/$$/$herschel.esac.esa.int$/$twiki$/$bin$/$view$/$Public$/$SummaryPointing.}
If there was a mis-pointing in the same direction for both the image and the spectrum, the unidentified FIR source could have been covered in the central spaxel of the spectral observations.  This
is because the unidentified FIR source is 6 $^\prime$$^\prime$ away from T~54 and  
the native PACS spaxel size is 9.4$\times$9.4 $^\prime$$^\prime$.  We note this possibility to be thorough.  However, we proceed by assuming all the [\ion{O}{1}] emission is from the T~54 system given that it also shows gas line emission in the MIR and leave it to future work to explore this further.

To conclude, the unidentified FIR source found by \citet{matra12} does not produce significant emission or contamination at the wavelengths of interest in this paper.  However, future works using data at 70~{\micron} and beyond should take the unidentified FIR source into account unless the emission can be spatially resolved.

\bibliographystyle{apjv2}

\end{document}